\documentclass{elsart}
\usepackage{graphicx}

\begin{document}

\begin{frontmatter}

\title{Quantum critical point in CuGeO$_3$ doped with magnetic impurities}

\author[address1]{S.V.Demishev\thanksref{thank1}},
\author[address2]{Y.Inagaki},
\author[address3]{M.M.Markina},
\author[address2]{H.Ohta},
\author[address2]{S.Okubo},
\author[address2]{Y.Oshima},
\author[address1]{A.A.Pronin},
\author[address1]{N.E.Sluchanko},
\author[address1]{N.A.Samarin},
\author[address1]{V.V.Glushkov}

\address[address1]{General Physics Institute of RAS, 38, Vavilov st., 119991 Moscow, Russia}

\address[address2]{Kobe University, 1-1 Rokkodai, Nada, Kobe 657-8501, Japan}

\address[address3]{Moscow State University, 119899 Moscow, Russia}

\thanks[thank1]{Corresponding author. E-mail: demis@lt.gpi.ru}

\begin{abstract}
Using high frequency (up to 450 GHz) ESR and low temperature
specific heat measurements we find that insertion of 1\% Fe and
2\% Co damps spin-Peierls and Neel transitions and for $T$$<$30K
may give rise to onset of a quantum critical behaviour characteristic
for a Griffiths phase.
\end{abstract}

\end{frontmatter}

Most of the available data for doped inorganic spin-Peierls compound
CuGeO$_{3}$ correspond to the limit of weak disorder
when density of states have a pseudogap \cite{p1,p2}.
In the oppposite case of a strong disorder the situation is expected to change
dramatically: the ground state of CuGeO$_{3}$
should be a Griffiths phase (GP) which thermodynamic properties
are controlled by relatively rare spin clusters correlated more
strongly than average \cite{p5,p6}. In this case density of states becomes
gapless and diverges at $\epsilon$$=$0:
$\rho(\epsilon)$$\propto$$\left|\epsilon\right|^{-\alpha}$ \cite{p1}.
As a consequence the temperature dependences of magnetic
susceptibility $\chi$ and magnetic contribution to specific
heat $c_{m}$ for Cu$^{2+}$ spin-Peierls chains acquire the forms
\cite{p5,p6}
\begin{equation}
\chi \propto T^{-\alpha};   c_{m} \propto T^{1-\alpha}
\label{e1}
\end{equation}
where $\alpha$$<$1. As long as $T$$=$0 becomes singular
or critical point, the aforementioned situation may be described
in terms of quantum critical (QC) behaviour \cite{p5}.

The experimental information about feasibility of the QC
point in CuGeO$_{3}$ is very limited
\cite{p7}. The aim of the present work consists in providing possible
evidence of onset of the QC point in CuGeO$_{3}$ doped with magnetic
impurities Co and Fe.

\begin{figure}[b]
\begin{center}\leavevmode
\includegraphics[width=0.95\linewidth]{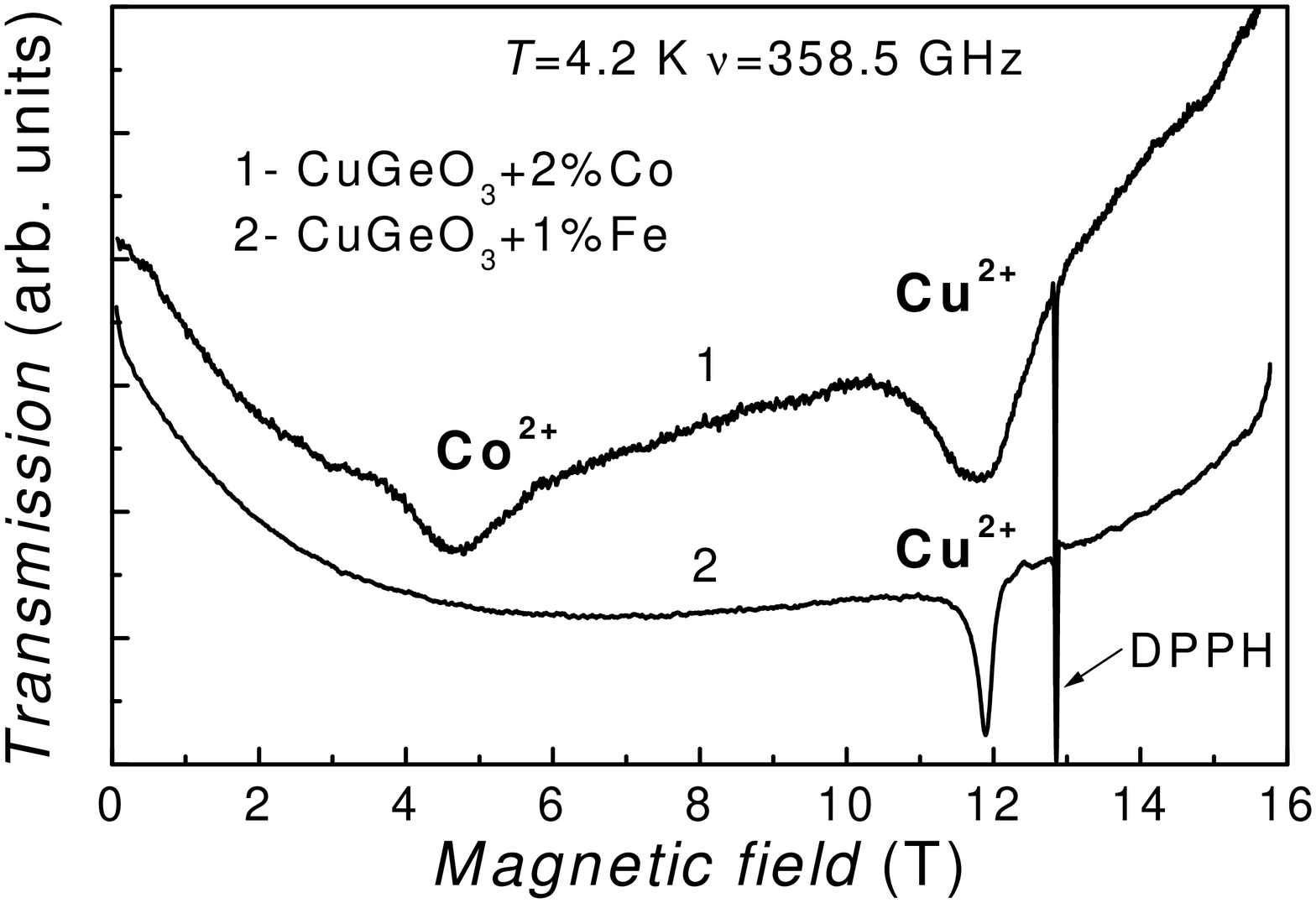}
\caption{Magnetoabsorption spectra for the samples studied.}
\end{center}
\end{figure}

Single crystals of CuGeO$_{3}$ doped with 1\% of Fe (S=1) and 2\%
of Co (S=3/2) impurities have been grown using self-flux method.
The quality of crystals was controlled by X-ray and Raman scattering
data \cite{p8}. Magnetic properties of the samples
in the temperature range 1.7-300K have been studied by
high (up to 450 GHz) frequency ESR technique \cite{p2}
for $\bf{B}$$\parallel$$\bf{a}$
geometry. Specific heat for the temperature
interval 6-20K was studied with the help of low temperature
small sample relaxation calorimeter.

Doping of CuGeO$_{3}$ with magnetic impurities causes
(i) rapid damping of the spin-Peierls
transition and (ii)
onset of a strong paramagnetic background \cite{p9}. Therefore the
deriving of the effects of doping on Cu$^{2+}$ S=1/2 chains
in convenient magnetisation measurements
requires \textit{ad hoc} assumptions about background behaviour \cite{p9}.
This difficulty overcomes in ESR experiment, where
magnetic impurity in CuGeO$_{3}$ matrix and Cu$^{2+}$ chains modified
by impurity gives rise to a different absorption lines \cite{p10}.

Typical magnetoabsorption spectra for CuGeO$_{3}$:Co and CuGeO$_{3}$:Fe
at helium temperatures are shown in Fig.1. For the Co-doped
sample spectrum is formed by two broad lines which becomes resolved
for frequencies $\nu$$>$100 GHz. The spectrum
of Fe-doped sample is characterised by a single resonance. The
resonant fields $B_{res}$ for all modes presented in Fig.1
are found to vary linearly with frequency $B_{res}$$\sim$$\nu$.
This result suggests that the observed magnetoabsorption features
correspond to different ESR modes rather than coexistence of
the ESR and antiferromagnetic resonance \cite{p1}.

The analysis of the \textit{g}-factor values together with the frequency
and temperature dependences of the linewidth \cite{p2} have allowed
to attribute the high field resonances with \textit{g}$\approx$2.15
to ESR on disordered Cu$^{2+}$ chains and the low field resonance
with \textit{g}$\approx$4.7 for CuGeO$_{3}$:Co to ESR on Co$^{2+}$ impurity
in CuGeO$_{3}$ matrix (Fig.1). A
strong \textit{g}-factor renormalisation for the Co-doped sample may
reflect formation of the impurity spin clusters \cite{p10}.

As long as the integrated intensity of the ESR line in the region
of linear response is proportional to magnetic susceptibility
it was possible to subtract contributions of Cu$^{2+}$ chains
and Co$^{2+}$ impurities unambiguously
and find $\chi(T)$ dependence for both contributions from the spectra
taken at different temperatures (Fig.2). The correctness of the $\chi(T)$
evaluation from ESR spectra has been checked by measurements of the static
susceptibility for Co-doped sample
by vibrating sample magnetometer. We found that sum of $\chi(T)$ for Cu$^{2+}$
and Co$^{2+}$  resonances represents the integral static susceptibility
of the sample within the error of less than 5\%.

Quantitative analysis of $\chi(T)$ data for doped
Cu$^{2+}$ chains indicates that Eq.(1) holds for $T$$<$30K
(curves 1 and 2 in Fig.2) and the power law
provides reasonable description of experimental data.
The best fit was obtained using index values $\alpha=$0.36$\pm$0.03
and $\alpha=$0.69$\pm$0.04 for Fe-doped and Co-doped crystals
respectively. In contrast to the magnetic properties
of Cu$^{2+}$ chains, the susceptibility for Co$^{2+}$ impurity is better
fitted by Curie-Weiss law,
\ensuremath{\chi}\ensuremath{\sim}1/(\textit{T}-\ensuremath{\Theta}),
with characteristic temperature $\Theta=$-2.8K
(Fig.2, curve 3) corresponding to antiferromagnetically interacting
Co$^{2+}$ impurities. For doped Cu$^{2+}$ chains it is visible that for magnetic
impurities studied spin-Peierls and Neel transitions are already damped.

\begin{figure}[t]
\begin{center}\leavevmode
\includegraphics[width=0.95\linewidth]{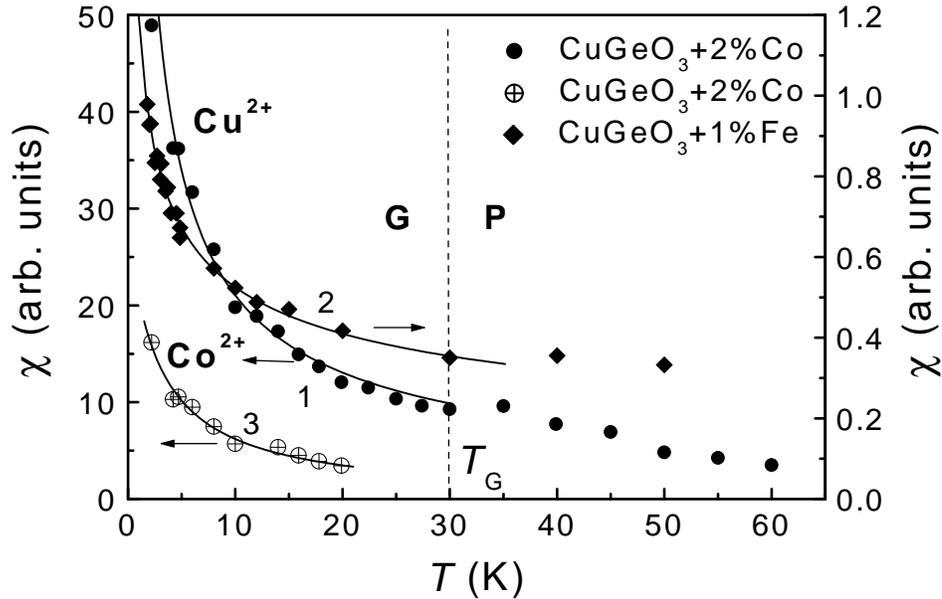}
\caption{Temperature dependence of magnetic susceptibility for doped
Cu$^{2+}$ chains and Co$^{2+}$ impurity. $\chi$$(T)$ data
suggest a transion from paramagnetic (P) to Griffiths (G) phase
at $T$$=$30K (see text for details).}
\end{center}
\end{figure}

The observation of the power asymptotics of $\chi(T)$
in the absence of any type of long range magnetic order for the wide range
1.7$<$$T$$<$30K (where temperature varies
more than 15 times) agrees with the theoretical predictions \cite{p5,p6}
for the QC point. An additional argument in favour of proposed
interpretation follow from specific heat data for CuGeO$_{3}$:Fe.
It was found \cite{p2} that magnetic contribution $c_{m}$ follows power
law with the index $\alpha=$0.37$\pm$0.03 which coincide
within experimental error with the value obtained from magnetic
susceptibility.

In conclusion, we provide possible experimental evidence that
doping of CuGeO$_{3}$ with magnetic impurities Co and Fe induces
a strong disorder limit and low temperature $T<$30K
ground state may become a GP. A magnetic structure of the clusters in GP,
including the possibility of the predicted in \cite{p11} random dimer phase,
is a subject of future investigations.

Authors acknowledge support from the INTAS project 00-807, programmes
''Physics of Nanostuctures'' and ''Integration'' of RAS and Venture Business
Laboratory of Kobe University.

%
%


\begin{thebibliography}{99}
\bibitem{p1} M.Mostovoy et al., Phys. Rev. B {\bf 58} (1998) 8190.
\bibitem{p2} S.V.Demishev et al., cond-mat/0110177.
\bibitem{p5} A.Rosch, in \textit{Abstracts of LT22, Helsinki} (1999) 389.
\bibitem{p6} D.Fisher, Phys. Rev. B {\bf 50} (1994) 3799.
\bibitem{p7} K.Manabe et al., Phys. Rev. B {\bf 58} (1998) R575.
\bibitem{p8} S.V.Demishev et al., JETP Letters {\bf 73} (2001) 31.
\bibitem{p9} P.E.Anderson et al., Phys. Rev. B {\bf 56} (1997) 11014.
\bibitem{p10} V.N.Glazkov et al., J.Phys.: Cond.
Mat. {\bf 10} (1998) 7879.
\bibitem{p11} H.Fabrizio, R.Mellin, Phys. Rev. Lett. {\bf 78} (1997) 3382.
\end{thebibliography}
\end{document}